\documentclass[eng
               ]{jetp}
\twocolumn
\begin{document}
\English
\title{Test of hybrid metric-Palatini f(R)-gravity in binary pulsars}
\author{N.~A.}{Avdeev}
\email{naavdeev1995@mail.ru}
\affiliation{Sternberg Astronomical Institute, Lomonosov Moscow State University, Universitetsky Prospekt, 13, Moscow, Russia}
\affiliation{Department of Astrophysics and Stellar Astronomy, Lomonosov Moscow State University, Leninskie Gory, 1/2, Moscow, Russia}

\author{P.~I.}{Dyadina}
\email{guldur.anwo@gmail.com}
\affiliation{Sternberg Astronomical Institute, Lomonosov Moscow State University, Universitetsky Prospekt, 13, Moscow, Russia}

\author{S.~P.}{Labazova}
\email{sp.labazova@physics.msu.ru}
\affiliation{Faculty of Mechanics and Mathematics, Lomonosov Moscow State University, Leninskie Gory, 1, Moscow 119991, Russia}

\rtitle{Test of hybrid metric-Palatini f(R)-gravity in binary pulsars}
\rauthor{N.~A.~Avdeev, P.~I.~Dyadina, S.~P.~Labazova}
\abstract
{
We developed the parameterized post-Keplerian formalism for hybrid metric-Palatini $f(R)$-gravity. We obtained analytical expressions in the generel eccentric case for four PPK parameters: $\dot\omega$, $\dot P_{\rm b}$, $r$ and $s$. Using observational data of PSR J0737-3039 and PSR J1903+0327 we imposed restrictions on the parameters of hybrid f(R)-gravity and showed that this theory is not ruled out by the observations in strong field regime. In addition we obtained predictions for masses of systems components and found that considered astrophysical objects will be heavier than in GR.
}

\date{March, 2020}

\maketitle

\section{Introduction}
General relativity (GR) is extremely successful theory which describes huge amounts of gravitational phenomena on a vast range of scales and gravitational regimes. This theory has  satisfyingly passed many tests at different scales~\cite{will14}. One of the last and, perhaps, the main achievements of GR is the prediction of gravitational waves, whose existence was registered by the LIGO detector~\cite{abbott}.

However, despite of this success, increased attention is paid to theories of gravity extending GR. The main reason for the growing interest in modified theories is inability to fully describe the observed late-time accelerated expansion of the Universe in the frameworks of GR only. Other reason for studying alternative theories of gravity is such puzzle of modern physics as dark matter~\cite{DM1, DM2}. Not everyone supports the idea of the need to search for new unknown particles. Some theories of gravity are created in order to provide a purely geometric description of the phenomena attributed to dark matter. One of such theories, which suggests a unified approach to the problems of dark energy and dark matter, is hybrid metric-Palatini f(R)-gravity~\cite{hybrid1,hybrid2}.

Hybrid f(R)-gravity belongs to the large family of f(R)-theories~\cite{fr1, fr2,fr3}.  The f(R)-gravity is one of the simplest ways to extend the Einstein-Hilbert action. All such theories are constructed by generalizing the gravitational part of the action as an arbitrary function of the scalar curvature $R$. The f(R)-gravity is successfully applied for description of inflation in early Universe~\cite{starobinsky}. Also such theories provide a beautiful explanation of late-time cosmic acceleration. Moreover, the accelerated expansion of the Universe is natural consequence of f(R)-theories. In addition, f(R)-gravity is attractive as an alternative to the $\Lambda$CDM model, since it allows to simultaneously describe early-time inflation and late-time cosmic acceleration~\cite{odintsov1, odintsov2, odintsov3, odintsov4, odintsov5, odintsov6, odintsov7,  saez}. Furthermore, f(R)-models are in good agreement with cosmological observational data and are almost indistinguishable from $\Lambda$CDM~\cite{odintsov8}.
	
	There are two main classes in f(R)-gravity: the metric one and the Palatini one. Each class is determined by the method of obtaining field equations. In the metric f(R)-model $g_{\mu\nu}$ is the only dynamical variable. In the Palatini f(R)-gravity  the Riemann curvature tensor is defined by  the metric and the independent affine connection. Thus, the metric approach provides fourth order differential field equations, while in the Palatini method these equations are of the second order~\cite{capo1,capo2}.
	
    	In many papers, the question of equivalence of two approaches has been studied in detail. In ~\cite{palamet, palamet1}  it was shown that the metric formalism is not equivalent to the Palatini (first-order) formalism. More general proof was given in the earlier work~\cite{ferraris, ferraris2}. It is clear that a generic non-linear matter-gravity theory may lead also to non-equivalent gravitational physics, depending on the choice of either metric or Palatini description. However, in the work~\cite{odinodin} authors prove that the Palatini formalism lead to accelerating universe in the same way as it happens in the metric formalism in the case of non-linear gravity-matter systems  with the Lagrangian $L=\sqrt{-g}(R+ f (R) L_{\rm d}) + k^2 L_{\rm mat}(\Psi)$, where $R$ is the curvature, $L_d$ is a scalar field Lagrangian and $L_{\rm mat}(\Psi)$ represents a matter Lagrangian; $k^2=8\pi G$. Then Big Rip singularities appear in the case of effective phantom models in vacuum universe, while quintessence models contain Big Bang like singularities. It was also shown that in the case of radiation models, where a radiation-like fluid is additionally considered, both Big Rip and Big Bang singularities appear.
	
	However,  both metric and Palatini approaches have several shortcomings. One of the fundamental disadvantages of metric f(R)-gravity is problems with passing standard tests in the Solar System~\cite{chiba, olmo1, olmo2}. Nevertheless, a limited class of viable models in the metric approach exists and was studied in detail in the papers~\cite{odintsov4, odintsov7, odintsov8}. Their viability is achieved through the chameleon mechanism~\cite{khoury, khoury1, hu, odintsov3}.
	
	On the other hand, the Palatini $f(\mathfrak{R})$-model can successfully describe both the late-time cosmic acceleration and Solar System. But despite this attractive property, Palatini $f(\mathfrak{R})$-models  lead to microscopic matter instabilities and to unacceptable features in the evolution patterns of cosmological perturbations~\cite{koivisto1, koivisto2}.
	
	Recently, a new class of f(R)-theories has been proposed~\cite{hybrid1}. It unites the advantages of both metric and Palatini f(R)-theories but lacks their shortcomings. This approach was called hybrid f(R)-theory. Such theories includes the metric part and the Palatini part. There is only one viable variant of hybrid f(R)-gravity, when the metric part is represented by the usual scalar curvature $R$, while the Palatini part is the general function of Palatini curvature $\mathfrak{R}$~\cite{hybrid2}. This hybrid metric-Palatini theory of gravity can be represented as dynamically equivalent scalar-tensor model. In this case, scalar field is long-range and plays an active role in cosmology. Also it provides the description of galactic dynamics. Moreover, the existence of such a scalar field is consistent with experiments in local systems even if the scalar field is very light. 
	
	The implication of hybrid f(R)-gravity was studied in many works. For example, static Einstein Universe was investigated in the work~\cite{bohmer}. Also, different cosmological models were studied in papers~\cite{lima, lea}. The accelerated expansion of the Universe was described in the work~\cite{capo3}. Moreover, the hybrid f(R)-gravity was investigated on astrophysical scales from stars to galaxy clusters. It was shown, that the virial mass discrepancy in clusters of galaxies can be explained via the geometric terms appearing in the generalized virial theorem~\cite{capo4}. The hybrid f(R)-gravity also allows to explain the rotational velocities of test particles gravitating around galaxies. This approach allows to avoid introducing of a huge amount of dark matter~\cite{capo5}. Besides,  physical properties of neutron, Bose-Einstein condensate and quark stars were considered~\cite{stars}. Furthermore, asymptotically anti-de Sitter wormhole solutions that satisfy the null energy condition for the whole spacetime were obtained\cite{capo6}. Also, stability of Kerr black holes in generalized hybrid metric-Palatini gravity was considered~\cite{2003}. In a recent work~\cite{meppn} a complete post-Newtonian (PPN) analysis was performed. The analytical expressions for $\gamma$ and $\beta$ parameters were obtained and it was proved that other 8 PPN parameters identically equal to zero. It was shown that the light scalar field in hybrid f(R)-gravity does not contradict the experimental data based not  on all set of PPN parameters. Moreover, hybrid f(R)-gravity was tested on the binary pulsars observational data~\cite{memnras}. In addition, the change of the orbital period  due to gravitational radiation was obtained in the quasicircular case. There are only scalar and tensor quadrupole terms in hybrid f(R)-gravity. Also for the first time the restriction on the scalar field  mass in hybrid f(R)-gravity was found~\cite{memnras}.

	Previously, the main attention was paid to the research of hybrid f(R)-gravity in cosmology and in the weak field limit. Only a couple of papers~\cite{memnras, stars, 2003} are devoted to the investigation of the effects  in the strong field regime. In this paper, we study manifestations of the hybrid f(R)-theory in the strong field of binary pulsars. For this aim we use parametrized post-Keplerian formalism (PPK)~\cite{DD, DT}.
	
Originally parametrized post-Keplerian formalism was developed to obtain dynamical information from the pulsar timing data in a theory-independent way. This dynamical information is encoded in a certain number of fitted post-Keplerian parameters. According to this formalism any modified theory of gravity can be described by 19 PPK parameters, which are functions of Keplerian parameters and inertial masses of the pulsar and its companion. Experimental values of eight PPK parameters can be obtained from pulsar timing, and eleven from the shape of the incoming pulses, and all 19 parameters can be measured independently. Thus, the PPK formalism becomes the powerful instrument for testing modified theories of gravity in the strong field regime, which realized in binary pulsars~\cite{DD, DT}.

	It is important to emphasize that strong field regime in binary pulsars is not so strong as in black holes, for example. Actually, using the term {\it strong field regime} in this context, we rather mean that in such systems the field is stronger than in the Solar System~\cite{DT}.  Indeed, since the pulsar is a rapidly rotating neutron star, on the surface of such object, the gravitational field is $ \left(\frac{GM}{c^2R} \right)_{\rm PSR} \sim0.2 $, while for the Sun it's only $ \left(\frac{GM}{c^2R} \right)_{\odot} \sim10^{- 6} $. In addition, the axis of the pulsar magnetic field  is shifted relatively the axis of its rotation. It causes modulation of periodic signals arriving on Earth. Due to the high stability of these pulses, relativistic effects of orbital motion can be observed, in particular the emission of gravitational waves.  It turns out that a strong gravitational field combined with high stability of the arrival of pulses makes binary pulsars a unique laboratory for the study and testing of modified gravity theories.
	
	The main aim of this article is test of hybrid f(R)-gravity using parametrized post-Keplerian formalism. In particular, we are going to obtain analytical expressions for four post-Keplerian parameters: $\dot\omega$ is change of periastron longitude, $\dot P_{\rm b}$ is orbital period decay, $r$ and $s$ are range and shape of Shapiro time delay respectively. Then using these expressions and  observational data of binary  pulsars PSR J0737-3039 and PSR J1903+0327 we will restrict the parameters of hybrid f(R)-gravity.
	
		The structure of the paper is the following. In section~\ref{sec1} we consider the action and the field equations of the hybrid metric-Palatini theory in a general form and in a scalar-tensor representation.  In section~\ref{sec2}, we discuss PPK formalism and obtain the analytical expressions for PPK parameters. Further, in section~\ref{sec3} we impose restrictions on the hybrid f(R)-gravity using the observational data of PSR J0737-3039 and PSR J1903+0327. We conclude in section~\ref{sec:conclusions} with a summary and discussion.
	
	Throughout this paper the Greek indices $(\mu, \nu,...)$ run over $0, 1, 2, 3$ and the signature is  $(-,+,+,+)$. All calculations are performed in the CGS system. The Jordan frame is used.

\section{Hybrid f(R)-gravity}\label{sec1}

The action of hybrid metric-Palatini f(R)-gravity consists of Hilbert-Einstein
term and an arbitrary function of the Palatini curvature~\cite{hybrid1,hybrid2}:
	\begin{gather}\label{act}
	S=\frac{c^4}{2k^2}\int d^4x\sqrt{-g}\left[R+f(\mathfrak{R})\right]+S_m,
	 \end{gather}
	where $c$ is the speed of light, $k^2=8\pi G$, $G$ is Newtonian gravitational constant, $R$ and $\mathfrak{R}=g^{\mu\nu}\mathfrak{R}_{\mu\nu}$ are the metric and Palatini curvatures respectively, $g$ is the metric determinant, $S_m$ is the matter action. Here the Palatini curvature $\mathfrak{R}$ is defined as a function of $g_{\mu\nu}$ and the independent connection $\hat\Gamma^\alpha_{\mu\nu}$:
\begin{equation}\label{re}
	\mathfrak{R}=g^{\mu\nu}\mathfrak{R}_{\mu\nu}=g^{\mu\nu} \bigl(\hat\Gamma^\alpha_{\mu\nu,\alpha}-\hat\Gamma^\alpha_{\mu\alpha,\nu}+\hat\Gamma^\alpha_{\alpha\lambda}\hat\Gamma^\lambda_{\mu\nu}-\hat\Gamma^\alpha_{\mu\lambda}\hat\Gamma^\lambda_{\alpha\nu} \bigr).
	 \end{equation}
	
	As in the case of the metric and Palatini f(R)-theories, the action \eqref{act} can be expressed in terms of a scalar field (for details see~\cite{hybrid1,hybrid2}):
	\begin{gather}\label{stact1}
	\begin{split}
	S=&\frac{c^4}{2k^2}\int d^4x\sqrt{-g}  \biggl[(1 + \phi)R + \frac{3}{2\phi}\partial_\mu \phi \partial^\mu \phi - V(\phi)  \biggr]\\
	&+S_m,
	\end{split}
	 \end{gather}
	where $\phi$ is a scalar field and $V(\phi)$ is a scalar field potential. This is the action of a non-minimally coupled scalar field with a non-canonical kinetic term.  Then the field equations become~\cite{hybrid1,hybrid2}:
	\begin{wide}
	\begin{eqnarray}
 	&&(1+\phi)R_{\mu\nu}=\frac{k^2}{c^4}\left(T_{\mu\nu}-\frac{1}{2}g_{\mu\nu}T\right)-\frac{3}{2\phi}\partial_\mu\phi\partial_\nu\phi +\frac{1}{2}g_{\mu\nu}  \biggl[V(\phi)+\nabla_\alpha\nabla^\alpha\phi  \biggr]+\nabla_\mu\nabla_\nu\phi,\label{feh}\\
 	&&\nabla_\mu\nabla^\mu\phi-\frac{1}{2\phi}\partial_\mu\phi\partial^\mu\phi-\frac{\phi[2V(\phi)-(1+\phi)V_\phi]}{3}=-\frac{k^2}{3c^4}\phi T_{\rm m},\label{fephi}
	 \end{eqnarray}
	 \end{wide}
where $T_{\mu\nu}$ and $T_{\rm m}$ are the energy-momentum tensor and its trace respectively.

\section{PPK formalism}\label{sec2}
The discovery of the system PSR B1913+16 in 1974~\cite{Hulse1975} provided  new opportunities for testing modified theories of gravity. This system is the first discovered binary pulsar. To analyze the pulsar timing data obtained from such systems, it was necessary to create a formalism that would allow extract information about the system in a theory-independent way. In 1986, Damour and Deruelle developed such formalism which allow to describe all effects up to the order $ \left(\frac{v^2}{c^2}\right) $ regardless of the gravitational theory~\cite{DD}. It was called the parametrized post-Keplerian formalism. Later, application of this formalism to test modified theories of gravity was described in the work~\cite{DT}. PPK formalism allows to test theories of gravity in the strong field limit similarly to parametrized post-Newtonian formalism~\cite{will, will14, will2}. The main idea of PPK formalism is a description of orbital effects in  the following form~\cite{DT}:
	\begin{gather}
		\label{tb-t0}
		t_{\rm b}-t_0=F\left[T;\{p^{\rm K}\};\{p^{\rm PK}\};\{q^{\rm PK}\}\right],
	\end{gather}
	where $t_{\rm b}$ denotes the solar-system barycentric (infinite-frequency) arrival time, $T$ is the pulsar proper time (corrected
for aberration). There are three sets of parameters~\cite{DT}:
\begin{equation}\label{kep param}
	\{p^{\rm K}\}=\{P_{\rm b},T_0,e_0,\omega_0,x_0\} 
\end{equation}
is the set of Keplerian parameters, where $P_{\rm b}$ is an orbital period, $e$ is an eccentricity, $\omega$ is an argument of periastron, $x$ is a projected semimajor axis of a pulsar's orbit;
\begin{equation}\label{postkep param}
	\{p^{\rm PK}\}=\{\dot \omega,\gamma,\dot{P_{\rm b}},r,s,\delta_\theta,\dot{e},\dot{x}\}
	\end{equation}
is the set of separately measurable post-Keplerian parameters, where $\gamma$ is a parameter of Einstein time delay, $\delta_\theta$ is a dimensionless parameter quantifying relativistic deformations of orbit, $\dot e$ is a secular change of eccentricity, $\dot x$ is a secular drift of projected semimajor axis;
and
\begin{equation}\label{postkep param q}
	\{q^{\rm PK}\}=\{\delta_r,A,B,D\}
	\end{equation}
	denotes the set of not separately measurable post-Keplerian parameters, where $\delta_r$ is a dimensionless parameter quantifying relativistic deformations of orbit connected with $\delta_\theta$: $A, B$ are parameters of abberation, $D$ is Doppler factor~\cite{DT}.

But in this work we take into consideration only four PPK parameters {$\dot\omega, \dot P_{\rm b}, r, s$}. Now let's discuss each of them in detail.

\subsection{Periastron advance $\dot \omega$}
 Start point of our consideration is obtaining of  Lagrangian describing the orbital motion. For this aim we use the method of Einstein, Infeld and Hoffmann (EIH)~\cite{EIH, will}. 
 
 Before proceeding to obtain the orbital Lagrangian we expand the scalar $\phi$ and tensor $g_{\mu\nu}$ fields of hybrid f(R)-gravity as
	\begin{equation}\label{decompos}
	\phi=\phi_0+\varphi,\qquad\ g_{\mu\nu}= \eta_{\mu\nu}+h_{\mu\nu},
	 \end{equation}
	where $\phi_0$ is the asymptotic background value of the scalar field far away from the source, $\eta_{\mu\nu}$ is the Minkowski background, $h_{\mu\nu}$ and $\varphi$ are the small perturbations of tensor and scalar fields respectively. In the general case $\phi_0$ is not a constant but the function of time $\phi(t)$. However this dependence can be neglected whenever its characteristic time scale is very long compared with the dynamical time scale associated with the local system itself. Thus, $\phi_0$ is taken as a constant. 
	
	The scalar potential $V(\phi)$ could be expanded in a Taylor series around the background value of scalar field $\phi_0$ like
	\begin{equation}\label{V}
	V(\phi)=V_0+V'\varphi+\frac{V''\varphi^2}{2!}+\frac{V'''\varphi^3}{3!}...
	 \end{equation}
	hence its derivative with respect to $\varphi$ will take the form $V_\phi=V'+V''\varphi+V'''\varphi^2/2$. Then the scalar field mass can be expressed as $m_\varphi^2=[2V_0-V'-(1+\phi_0)\phi_0V'']/3$~\cite{meppn}. 
 
 Thus, the EIH Lagrangian takes the following form: 
	\begin{wide}
	\begin{eqnarray}
		\label{L_rel_hybrid}
		L^{rel}\left(\mathbf{R},\mathbf{V}\right)\equiv&&\mu^{-1}L_O^{rel}\left(\mathbf{R},\mathbf{V}\right)=\dfrac{1}{2}\mathbf{V}^2+\dfrac{GM}{R}\dfrac{1}{\left(1+\phi_0\right)}\left(1-\dfrac{\phi_0}{3}e^{-m_\varphi R}\right)+\dfrac{1}{8}\left(1-3\nu\right)\dfrac{\mathbf{V}^4}{c^2}\nonumber\\
		&+&\dfrac{GM}{2Rc^2}\Bigg[\dfrac{1}{\left(1+\phi_0\right)}\bigg(3+\nu+\dfrac{\phi_0}{3}e^{-m_\varphi R}\left(1-\nu\right)\bigg)\mathbf{V}^2+\frac{1}{\left(1+\phi_0\right)}\left(1-\dfrac{\phi_0}{3}e^{-m_\varphi R}\right)\nu\left(\mathbf{N}\cdot\mathbf{V}\right)^2
		\\ \nonumber
		&-&\dfrac{GM}{R}\frac{9e^{2m_\varphi R}-\phi_0\left(6e^{m_\varphi R}+1\right)}{\left(3e^{m_\varphi R}-\phi_0\right)^2}\Bigg],
	 \end{eqnarray}
	 \end{wide}
where $\mathbf{R}\equiv\mathbf{r_1}-\mathbf{r_2}$, $\mathbf{V}=\mathbf{v_1}-\mathbf{v_2}$, $m, m'$ are pulsar's and companion's masses, $M = m + m'$ denotes total mass, $\mu =\frac{mm'}{m+m'}$ is effective mass, $\nu = \frac{\mu}{M}$, ${\bf N}$ is unit vector in direction of emission in the pulsar comoving frame.

Since Lagrangian \eqref{L_rel_hybrid} is invariant under time shifts and spatial rotations, there are four first integrals of motion: the energy of the system $E$ and angular momentum ${\bf J}$:
	\begin{gather}
	\begin{split}
		E=\mathbf{V}\cdot\cfrac{\partial L^{rel}}{\partial\mathbf{V}}-L^{rel},\qquad \mathbf{J}=\mathbf{R}\times\cfrac{\partial L^{rel}}{\partial\mathbf{V}}.
	\end{split}
	\end{gather}
Using expression  \eqref{L_rel_hybrid} it is possible to represent energy and angular momentum in the following forms:
	\begin{gather}
	\begin{split}
		\label{E}
		E=&\frac{1}{2}\mathbf{V}^2-\frac{GM}{R}\frac{1}{\left(1+\phi_0\right)}\left(1-\dfrac{\phi_0}{3}e^{-m_\varphi R}\right)+\frac{3}{8}\left(1-3\nu\right)\frac{\mathbf{V}^4}{c^2}
		\\
		+&\frac{GM}{2Rc^2}\Bigg[\frac{1}{\left(1+\phi_0\right)}\left(3+\nu+\dfrac{\phi_0}{3}e^{-m_\varphi R}\left(1-\nu\right)\right)\mathbf{V}^2
		\\
		+&\frac{\left(\mathbf{N}\cdot\mathbf{V}\right)^2}{\left(1+\phi_0\right)}\left(1-\dfrac{\phi_0}{3}e^{-m_\varphi R}\right)\nu\\
		+&\frac{GM}{R}\frac{9e^{2m_\varphi R}-\phi_0\left(6e^{m_\varphi R}+1\right)}{\left(3e^{m_\varphi R}-\phi_0\right)^2}\Bigg],
	\end{split}
	\end{gather}
	\begin{gather}
	\begin{split}
		\label{J}
		\mathbf{J}=&\mathbf{R}\times\mathbf{V}\Bigg[1+\frac{1}{2}\left(1-3\nu\right)\frac{\mathbf{V}^2}{c^2}
		\\
		&+\frac{GM}{Rc^2}\frac{1}{\left(1+\phi_0\right)}\left(3+\nu+\dfrac{\phi_0}{3}e^{-m_\varphi R}\left(1-\nu\right)\right)\Bigg].
	\end{split}
	\end{gather}

The existence of the first integral~\eqref{J} means that the system components move in the coordinate plane. So it's convenient to turn to the polar coordinates $\{R,\theta\}$: $R_x=R\cos{\theta}$, $R_y=R\sin{\theta}$ and $R_z=0$. Then, using the identities:
	\begin{gather}
	\begin{split}
		&\mathbf{V}^2=\left(\frac{dR}{dt}\right)^2+R^2\left(\frac{d\theta}{dt}\right)^2,
		\\
		&|\mathbf{R}\times\mathbf{V}|=R^2\frac{d\theta}{dt},
		\\
		&\left(\mathbf{N}\cdot\mathbf{V}\right)=\frac{dR}{dt},
	\end{split}
	\end{gather}
and neglecting all terms up to the order~$\left(\frac{v}{c}\right)^2$, we can represent the equations of relative motion in polar coordinates using expressions for the first integrals ~\eqref{E}-\eqref{J}:
	\begin{align}
		\label{dR/dt}
		\left(\frac{dR}{dt}\right)^2&=A+\frac{2B}{R}+\frac{C}{R^2}+\frac{D}{R^3},
		\\
		\label{dtheta/dt}
		\frac{d\theta}{dt}&=\frac{H}{R^2}+\frac{I}{R^3},
	\end{align}
where
	\begin{align}
		\label{A}
		A&=2E\left[1+\cfrac{3}{2}\left(3\nu-1\right)\cfrac{E}{c^2}\right],
		\end{align}
		\begin{align}\label{B}
		B&=GM\cfrac{1}{\left(1+\phi_0\right)}\Bigg\{\left[1-\dfrac{\phi_0}{3}e^{-m_\varphi R}\right]\\
		&+\left[7\nu-6-\dfrac{\phi_0}{3}e^{-m_\varphi R}\left(7\nu-2\right)\right]\cfrac{E}{c^2}\Bigg\},
		\end{align}
		\begin{align}\label{C}
		C&=-J^2\left[1+2\left(3\nu-1\right)\cfrac{E}{c^2}\right]\\
		&+\Bigg\{\cfrac{1-\dfrac{\phi_0}{3}e^{-m_\varphi R}}{\left(1+\phi_0\right)^2}\Big[5\nu-9-\dfrac{\phi_0}{3}e^{-m_\varphi R}\left(5\nu-1\right)\Big]
		\nonumber\\
		&-\frac{9e^{2m_\varphi R}-\phi_0\left(6e^{m_\varphi R}+1\right)}{\left(3e^{m_\varphi R}-\phi_0\right)^2}\Bigg\}\cfrac{G^2M^2}{c^2},
		\end{align}
		\begin{align}
		\label{D}
		D&=\cfrac{1}{\left(1+\phi_0\right)}\left[-3\nu\left(1-\dfrac{\phi_0}{3}e^{-m_\varphi R}\right)+8\right]\cfrac{GMJ^2}{c^2},
		\end{align}
		\begin{align}\label{H}
		H&=J\left[1+\left(3\nu-1\right)\cfrac{E}{c^2}\right],
		\end{align}
		\begin{align}\label{I}
		I&=\cfrac{1}{\left(1+\phi_0\right)}\left[2\nu\left(1-\dfrac{\phi_0}{3}e^{-m_\varphi R}\right)-4\right]\cfrac{GMJ}{c^2}.
	\end{align}

Follow the method described in~\cite{DD}, we solve these equations and obtain expression for the periastron advance:
 	\begin{align}
		\label{omega1}
		\dot{\omega}=n\left(K-1\right),
	\end{align}
where $n$ is average motion:
	\begin{align}
		\label{n(EJ)}
		n=&\frac{\left(-2E\right)^{3/2}\left(1+\phi_0\right)}{GM\left(1-\dfrac{\phi_0}{3}\right)}\Bigg\{1-\frac{1}{4\left(1-\dfrac{\phi_0}{3}\right)}\\
		&\times\biggl[\nu-15-\dfrac{\phi_0}{3}\left(\nu+1\right)\biggr]\frac{E}{c^2}-\frac{GM\phi_0}{2E\left(1+\phi_0\right)}m_\varphi\Bigg\},
	\end{align}
	\begin{align}
		\label{K}
		K=\frac{H_e}{n(a'')^2\sqrt{1-e_\theta^2}},
	\end{align}
where $e_{\theta}$ is an effective eccentricity:
    \begin{gather}
	\begin{split}
		\label{etheta}
		e_\theta^2=&1+\frac{2E\left(1+\phi_0\right)^2}{G^2M^2\left(1-\dfrac{\phi_0}{3}\right)^2}\\
		&\times\Biggl\{1+\Biggl[\frac{17}{2}\nu-\frac{7}{2}-\frac{4}{1-\dfrac{\phi_0}{3}}\biggl[2\nu+1\\
		&-\dfrac{\phi_0}{3}\left(2\nu-1\right)\biggr]\Biggr]\frac{E}{c^2}\Biggr\}\Biggl\{J^2+\Biggl[\frac{1-\dfrac{\phi_0}{3}}{\left(1+\phi_0\right)^2}\left[-7-\dfrac{\phi_0}{3}\right]
		\\
		&+\frac{9-7\phi_0}{9\left(1-\dfrac{\phi_0}{3}\right)^2}\Biggr]\frac{G^2M^2}{c^2}\Biggr\}+\frac{2\phi_0\left(1+\phi_0\right)J^2}{2GM\left(1-\dfrac{\phi_0}{3}\right)^2}m_\varphi,
	\end{split}
	\end{gather}
and $a''$ is an effective semimajor axis:
	\begin{gather}
	\begin{split}
		\label{a''new}
		a''=&-\frac{GM}{2E\left(1+\phi_0\right)}\left[1-\dfrac{\phi_0}{3}\right]\Bigg\{1-\left[1-3\nu\right]\frac{E}{2c^2}\Bigg\}
		\\
		&+\left(\frac{GM}{2E\left(1+\phi_0\right)}\right)^2\left[1-\dfrac{\phi_0}{3}\right]\frac{2\phi_0}{3}m_\varphi.
	\end{split}
	\end{gather}
	Here we also used the following notification:
	\begin{align}
		\label{He}
		H_e=&J\left[1+\left(3\nu-1\right)\frac{E}{c^2}\right].
	\end{align}
To obtain this result we considered the case of light scalar field $m_{\phi}R<<1$. The scalar field mass is responsible for dark energy effect. The manifestations of influence of this effect start from the distances much larger than the separation of components in binary pulsars. Therefore, we can neglect the terms of the form $~\frac{m_\varphi}{c^2}$, and use the Taylor expansion of the function $e^{-m_\varphi R}$.

As a result, we obtain the following expression for periastron advance:
 	\begin{wide}
	\begin{eqnarray}
		\label{omegadotfin}
		&&\dot{\omega}=\frac{\left(GM\right)^{2/3}\pi\left(2\pi\right)^{2/3}}{c^2P_{\rm b}^{5/3}\left(1-e^2\right)}\frac{\left(1+\phi_0\right)^{4/3}}{\left(1-\dfrac{\phi_0}{3}\right)^{4/3}}\left[\frac{1-\dfrac{\phi_0}{3}}{\left(1+\phi_0\right)^2}\left[7+\dfrac{\phi_0}{3}\right]-\frac{9-7\phi_0}{9\left(1-\dfrac{\phi_0}{3}\right)^2}\right]
		-\frac{\left(GM\right)^{1/3}4\phi_0\left(4^2\pi\right)^{1/3}}{3P_{\rm b}^{1/3}\left(1+\phi_0\right)^{1/3}\left(1-\dfrac{\phi_0}{3}\right)^{2/3}}m_\varphi. \nonumber\\
		&&
	 \end{eqnarray}
	 \end{wide}
	
\subsubsection{PPK parameters $s$ and $r$}

Now we turn to the Shapiro time delay. Firstly we consider the parameter $s$ which characterizes the shape of Shapiro time delay and is equal to the sine of the orbit inclination $\sin i$. 

A motion in binary system obeys the Kepler's third law:
\begin{equation}\label{Kepler} 
a^3(2\pi/P_{\rm b})^2=G^{\rm eff}m,
\end{equation} 
where $a$ is semi-major axis of relative orbit, $G^{\rm eff}$ is effective gravitational constant~\cite{meppn}:
	\begin{align}
		\label{Geff}
		G^{\rm eff}&=\frac{G}{\left(1+\phi_0\right)}\left(1-\frac{\phi_0}{3}e^{-m_\varphi R}\right).
	\end{align}
We considered the case of light scalar field $m_{\phi}R<<1$. Since based on Kepler's third law for hybrid f(R)-gravity \eqref{Geff} we can find semimajor axis $a$:
\begin{align}
	\label{a(n)}
	a=\left(\frac{GMP_{\rm b}^2}{4\pi^2\left(1+\phi_0\right)}\right)^{1/3}\left(1-\dfrac{\phi_0}{3}\right)^{1/3}.
	\end{align}
	
By definition the semimajor axis of pulsar orbit about the center of mass $a_1=\frac{m_2a}{M}\equiv\frac{cx}{\sin{i}}$, so using the expression for $a$~\eqref{a(n)}, we obtain:
	\begin{align}
		\label{s}
		s=\sin i=\frac{cxM}{m_2a}=\frac{cx\left(4\pi^2\right)^{1/3}M^{2/3}}{G^{1/3}P_{\rm b}^{2/3}m_2}\left(\frac{1+\phi_0}{1-\dfrac{\phi_0}{3}}\right)^{1/3}.
	\end{align}

The next parameter $ r $ describes the range of Shapiro time delay.  To express this parameter it is necessary to use the equation of photon motion along null geodesics:
	\begin{align}\label{geod}
		-1+h_{00}^{(2)}+\left(\delta_{ij}+h_{ij}^{(2)}\right)u^iu^j=0,
	\end{align}
where 
	\begin{align}
		\label{h002}
		h_{00}^{(2)}&=\frac{2}{c^2}G^{\rm eff}\frac{M}{R},
		\\
		\label{hij2}
		h_{ij}^{(2)}&=\frac{2}{c^2}G^{\rm eff}\frac{M}{R}\delta_{ij}
		\\
	\end{align}
	are perturbations of the metric up to the order $O(2)$~\cite{meppn}, $\delta_{ij}$ is the Kronecker delta. Also it is necessary to take into account that $u^{\mu}=d x^{\mu}_a/d \tau_a$ is four-velocity of the $a$-th particle, $d\tau=\sqrt{-ds^2}/c$, $ds^2=g_{\mu\nu}dx^{\mu}dx^{\nu}$ is an interval.
	
Thus \eqref{geod} takes the form:
	\begin{align}
		-1+\frac{2}{c^2}G^{\rm eff}\frac{m_2}{r}+\left(1+\frac{2}{c^2}G^{\rm eff}\gamma_{\rm PPN}\frac{m_2}{r}\right)|\mathbf{u}|^2=0,
	\end{align}
where
	\begin{align}
		\label{gammappn}
		\gamma_{PPN}&=\frac{1+\cfrac{\phi_0}{3}e^{-m_\varphi R}}{1-\cfrac{\phi_0}{3}e^{-m_\varphi R}}
	\end{align}
is the effective post-Newtonian parameter $\gamma$~\cite{meppn}.
	
If the photon was emitted at the point $\mathbf{x_e}$ in the direction of $\mathbf{n}$ at the time $t_e$, then its trajectory 
taking into account the post-Newtonian (PN) corrections $x^i_{\rm PN}(t)$ is described by the expression:
	\begin{align}
		\label{xi(t)}
		x^i(t)=x_e^i+n^i\left(t-t_e\right)+x^i_{\rm PN}(t).
	\end{align}
	
Using identity
	\begin{align}
		\label{u2}
		|\mathbf{u}|^2=1+2\left(\mathbf{n}\cdot\frac{d\mathbf{x}_{\rm PN}(t)}{dt}\right)=1+2\frac{dx_{\rm PN}^{\parallel}(t)}{dt}+O(c^4),
	\end{align}
we obtain
	\begin{align}
		\label{dxdt}
		\frac{dx_{\rm PN}^{\parallel}(t)}{dt}=-\frac{1}{c^2}G^{\rm eff}\left(1+\gamma_{\rm PPN}\right)\frac{m_2}{r}.
	\end{align}
Then the time of photon traveling from $\mathbf{x_e}$ to $\mathbf{x}$ and back equal to:
	\begin{align}
		\Delta t=\frac{2}{c}|\mathbf{x}-\mathbf{x_e}|-\frac{1}{c^3}\int\limits_{t_e}^tG^{\rm eff}_{PPN}\left(1+\gamma_{PPN}\right)\frac{m_2}{r}dt',
	\end{align}
and the parameter $ r $ from the Shapiro time delay, respectively equal~\cite{alsing, DT}
	\begin{align}
		\label{r}
		r=\frac{Gm_2}{c^3\left(1+\phi_0\right)}.
	\end{align}

\subsubsection{First derivative of orbital period $\dot P_{\rm b}$}
In~\cite{memnras} we considered first derivative of orbital period in the case of quasicircular orbit. But in this article we obtain expression for gravitational energy flux from binary pulsars in the general case of an eccentric orbit:
	\begin{align}
		\label{Egrav}
		\left<\dot E_{\rm grav}\right>=\left<\dot E_{\rm tensor}\right>+\left<\dot E_{\rm scalar}\right>,
	\end{align}
where tensor part is
\begin{align}\label{gravitational flux2} 
\left<\dot E_{\rm tensor}\right> =-\frac{G}{5c^5(1+\phi_0)} \left< \dddot{M}^{kl}\dddot{M}_{kl}-\frac{1}{3}(\dddot{M}^{kk})^2\right>,
\end{align}
\begin{align}\label{quadrupole moment} 
M_{ij} =\sum_a m^a(\phi)r_{i}^a(t)r_{j}^a(t)
\end{align}
is  quadrupole moment and scalar part is
\begin{align}\label{scalar_flux}
\left<\dot E_{\rm scalar}\right> =&&\frac{2c^5G\phi_0}{6(1+\phi_0)}\int dz_1dz_2J_1(z_1)J_2(z_2) \biggl\langle \frac{1}{c^6}\mathcal{\dot M}_0\mathcal{\dot M}_1\nonumber\\
&&+\frac{1}{6c^8}\biggl(2\mathcal{\ddot M}_1^k\mathcal{\ddot M}_2^k+\mathcal{\dot M}_0\mathcal{\dddot M}_3^{kk}+\mathcal{\dot M}_1\mathcal{\dddot M}_2^{kk}\biggr)\nonumber\\
&&+\frac{1}{60c^{10}}\biggl(2\mathcal{\dddot M}_2^{kl}\mathcal{\dddot M}_3^{kl}+\mathcal{\dddot M}_2^{kk}\mathcal{\dddot M}_3^{ll}\biggr)\nonumber\\
&&+\frac{1}{30c^{10}}\biggl(\mathcal{\ddot M}_1^{k}\mathcal{\ddddot M}_4^{kll}+\mathcal{\ddot M}_2^{k}\mathcal{\ddddot M}_3^{kll}\biggr)\biggr\rangle,
\end{align} 
where $J_1(z)$ is the first order Bessel function, $z=m_\varphi\sqrt{c^2(t-t')^2-|\mathbf{r}-\mathbf{r}'|^2}$ and
\begin{align}\label{mll} 
\mathcal{M}^L_l&=&\mathcal{M}^{i_1i_2...i_l}_l(t, r, z)=\sum_a\biggl(M_a(t-r/c)r^L_a(t-r/c)\nonumber\\
&&-u^{-(l+1)}(r,z)M_a(t-ru(r,z)/c)r^L_a(t-ru(r,z)/c)\biggr),\nonumber\\
\end{align}
where
\begin{align}\label{m} 
M_a(t)=&&m_a\biggl[1-\frac{v_a^2}{2c^2}-3\sum_{b\neq a}\frac{Gm_b}{r_{ab}(t)c^2(1+\phi_0)}\\
&&-\frac{G\phi_0}{c^2 (1+\phi_0)}\sum_{b\neq a}\frac{m_b}{r_{ab}(t)}e^{-m_\varphi R}\biggr].\nonumber\\
\end{align} 
Here $r^L_a(t)=r_a^{i_1}(t)r_a^{i_2}(t)...r_a^{i_l}(t)$,  $u(r, z)=\sqrt{1+(z/m_\varphi r)^2}$, $m_a$ and $v_a$  are mass and velocity of object $a$ respectively. Dots denote derivatives with respect to time.

Unlike the derivation of the expression for the first derivative of the orbital period shown in~\cite{memnras}, here we use the case of the eccentric orbit:
\begin{align}\label{r_eccentric} 
r=\frac{a(1-e^2)}{1+e\cos{\theta}}.\nonumber
\end{align} 
Obtaining this equation we also suggest that scalar field is very light. Thus calculating the energy flux due to the gravitational radiation and using the expression~\cite{memnras}:
\begin{equation}\label{energy_losses} 
\cfrac{\dot E}{E}=-\cfrac{2}{3}\cfrac{\dot P_{\rm b}}{P_{\rm b}},
\end{equation} 
we find:
\begin{wide}
\begin{align}\label{r_eccentric} 
\frac{\dot P_{\rm b}}{\dot P^{GR}_{\rm b}} = \frac{(1-\frac{\phi_0}{3})^{2/3}}{(1+\phi_0)^{5/3}}\Biggl[1+\frac{(19\phi_0-57)\phi_0}{15552(1-\frac{\phi_0}{3})^3(1+\frac{73}{24}e^2+\frac{37}{96}e^4)}\Biggl(\frac{1}{19}((15e^4+64e^2-12)(3-\phi_0)^2\nonumber\\
-\frac{8973}{152}\Biggl(\Biggl(\phi_0^2+\frac{5538\phi_0}{997}+9\Biggr)e^4 +\Biggl(\frac{62200\phi_0^2}{8973}+\frac{117520\phi_0}{2991}+\frac{62200}{997}\Biggr)e^2 + \frac{17440}{997} + \frac{34240\phi_0}{2991}+\frac{17440\phi_0^2}{8973}\Biggr)\Biggr)\Biggr].
\end{align}
\end{wide}

Thus the change of the orbital period in binary pulsars occurs due to energy loss to scalar and tensor radiation. The expression \eqref{r_eccentric} includes tensor and scalar parts. There is only quarupole contribution in the tensor sector. The tensor part coincides with the value of the orbital period decay predicted by GR up to the effective gravitational constant $G^{\rm eff}$ between components of system. The scalar sector includes PN corrections to the monopole term,  monopole-quadrupole and quadrupole contributions. The monopole term vanishes in the quasi-circular approximation~\cite{memnras} but in the eccentric case it survives, same as monopole-quadrupole contribution. It is also important to emphasize that hybrid f(R)-gravity does not predict the presence of scalar dipole radiation either in the quasi-circular or in the more general eccentric case.

\section{Observational limits}\label{sec3}
For the PPK test we used data from two binary systems with a pulsar: PSR J0737-3039~\cite{kramer} and PSR J1903+0327~\cite{freire}. First of them is system, which consists of two pulsars and the second is mixed system, which includes pulsar and the main sequence star. 

PSR J0737-3039 is the only known double binary pulsar.  The extraordinary closeness of the system components, small orbital period and the fact that
we see almost edge-on system allow to investigate the manifestation of relativistic effects with the highest available precision. Since the system PSR J0737-3039 components are pulsars  it is possible to measure semi-major axis of the orbit for each component. This fact leads to the possibility to obtain value of ratio:
\begin{gather}\label{R}
  \begin{split}
    \cfrac{a_2}{a_1} = \cfrac{m_1}{m_2} = R, 
   \end{split}
\end{gather}
i.e. the ratio of the masses can be measured directly. All the mentioned facts make PSR J0737-3039 good laboratory for testing modified theories of gravity~\cite{kramer}. It is important to note that we used this system to test the hybrid f(R)-theory in the case of a quasicircular orbit. Indeed, the eccentricity of the orbit is small in this system, however, taking into account its nonzero values we improve the accuracy of the restrictions imposed on the considered model.

	\begin{table}[h]
\caption{Parameters  PSR J0737-3039~\cite{kramer}}\label{0737}
\begin{tabular}{|l|l|l|}
\hline
Parameter & Physical & Experimental\\
 & meaning & value\\
\hline
$P_{\rm b} (day)$ & orbital period  & $ 0.10225156248(5)  $\\
\hline

$e$ & eccentricity & $ 0.0877775(9)   $\\
\hline

$x (s)$ & projected   & $ 1.415032(1)  $\\
& semimajor axis & \\
& of the pulsar orbit &\\
\hline

$\dot \omega(deg/yr)$ & secular advance  & $ 16.89947(68)$\\
&of the periastron&\\

 \hline
$\dot P_{\rm b}$ & secular change of  & $ -1.252(17) $\\
& the orbital period & $\times 10^{-12}$\\

\hline
$s$ & Shapiro ``shape''  & $0.99974_{-39}^{+16} $\\
 & delay parameter & \\
\hline
$r(\mu s)$ & Shapiro ``range''  & $ 6.21(33)  $\\
 & delay parameter & \\
\hline
$R=\cfrac{m_1}{m_2}$ & mass ratio & $ 1.0714(11)  $\\
\hline
\end{tabular}
\end {table}

It is necessary to emphasize that the observational value of the orbital period change $\dot P_{\rm b}$ can include various components which have the different nature: intrinsic and kinematic effects~\cite{DT, laza}. We are interested only in such systems where the dominant part of observable $\dot P_{\rm b}$ is the orbital period change due to the emission of gravitational waves. Thus we don't consider binary pulsars where effects of the mass loss and tidal torques are important in relation to the effect of gravitational radiation.

We used the method of testing models of gravity from the work~\cite{DT} and our additions. We plot the curves representing post-Keplerian parameters on the plane, in the $Y$-axis we put the allowed values of the companion masses $m_2$ and in the $X$-axis we put possible values of the pulsar $m_1$ masses. The region of all curves intersection within the measured accuracy displays the possible range of the pulsar's mass and its companion. For the appropriate gravity model all the curves of post-Keplerian parameters should ''meet'' at one point (taking into account the accuracy). However, if the curves diverge at some values of the model parameters, this 
means that at these values the theory does not work.

\begin{figure}[ht!]
		\center{a)\includegraphics[width=0.8\linewidth]{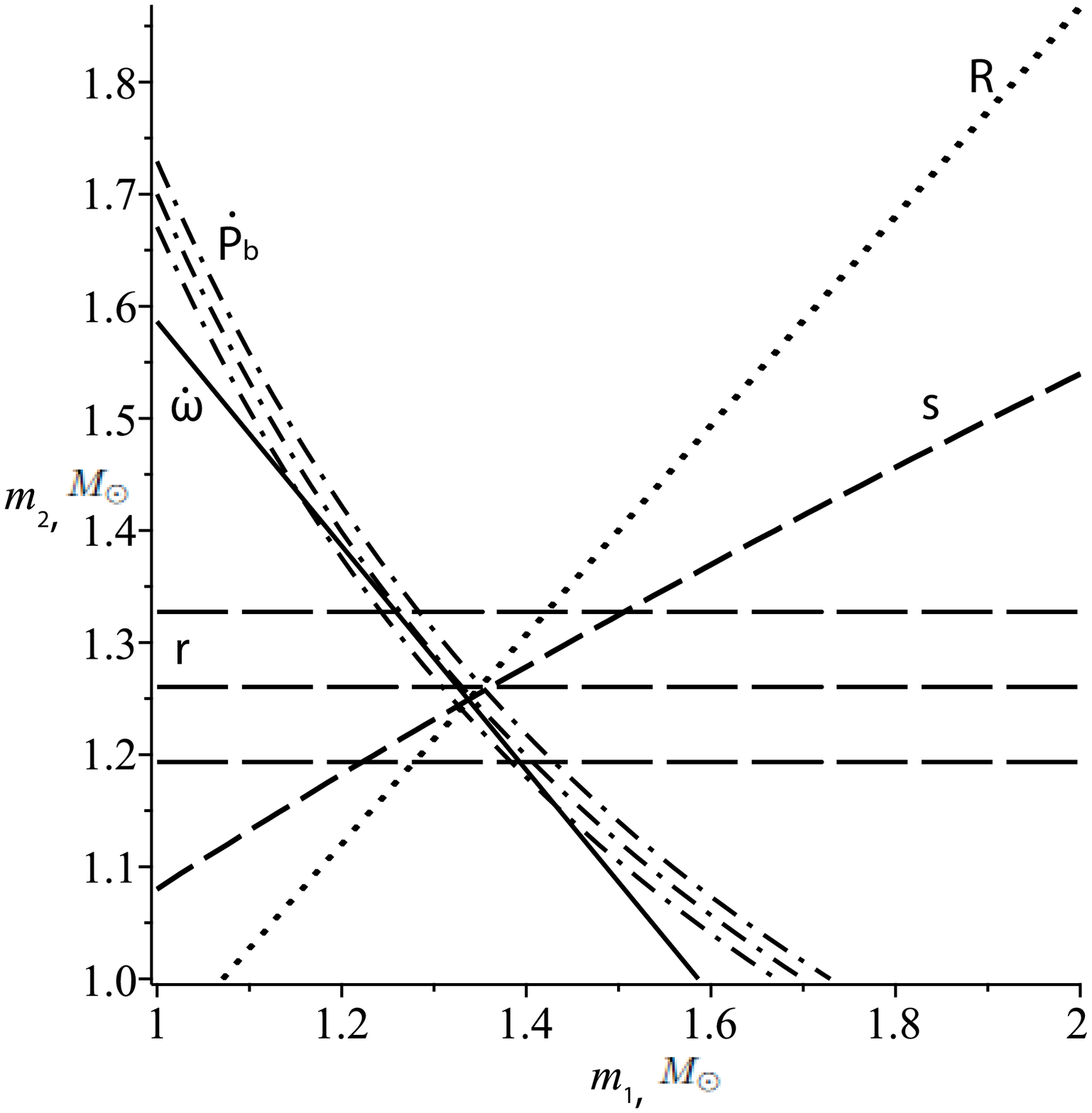}\label{pic GR 1}}
		{b)\includegraphics[width=0.8\linewidth]{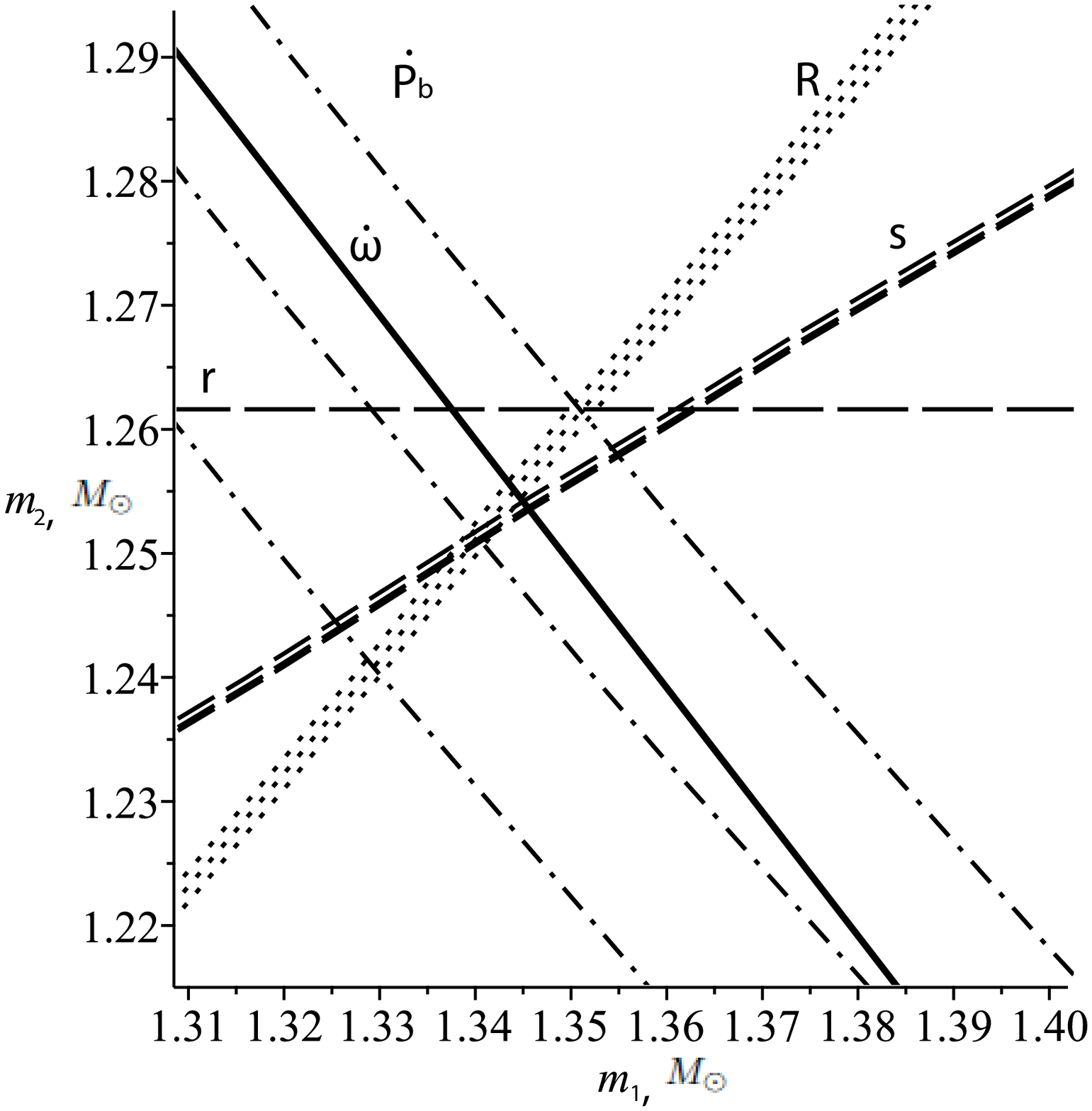}\label{pic GR 2}}
		\caption{Mass-mass diagrams for PSR J0737-3039.
The width of each curve represents $\pm1\sigma$ error bounds.
a) GR case; b) hybrid f(R)-gravity case at boundary values of model parameters.}
		\label{pic GR}
	\end{figure}

In this work we obtained the analytical expressions for four PPK parameters: $\dot \omega, \dot P_{\rm b}, r, s$. Then using observational data from PSR J0737-3039 (see Table \ref{0737}) we represented all these parameters on the fig. \ref{pic GR}. Changing parameters of hybrid f(R)-gravity $\phi_0$ and $m_\phi$ we find  values of these parameters where curves don't intersect within the measurement accuracy. Thus we impose restrictions on hybrid f(R)gravity: $\phi_0<0.001$ and $\frac{\phi_0 m_{\phi}}{(1+\phi)^{\frac{1}{3}}(1-\frac{\phi_0}{3})^{\frac{2}{3}}}<1.3\times 10^{-17}$. But in fact we can't impose restrictions on the scalar field mass, since we managed to restrict only the above combination. In addition, we obtained predictions for the masses of system components in the frameworks of hybrid f(R)-gravity: $1.3374 \ \  M_\odot<m_1<1.3440 \ \  M_\odot$ and $1.2482 \ \  M_\odot<m_2<1.2537 \ \  M_\odot$. At the same time, GR predicts the following mass values in this system: $1.3374 \ \  M_\odot<m_1<1.3388 \ \  M_\odot$ and $1.2482 \ \  M_\odot<m_2<1.2496 \ \  M_\odot$~\cite{kramer}. As we can see the presence of a scalar field increases the possible observed masses of pulsars.	

\begin{figure}[ht]
		\center{a)\includegraphics[width=0.8\linewidth]{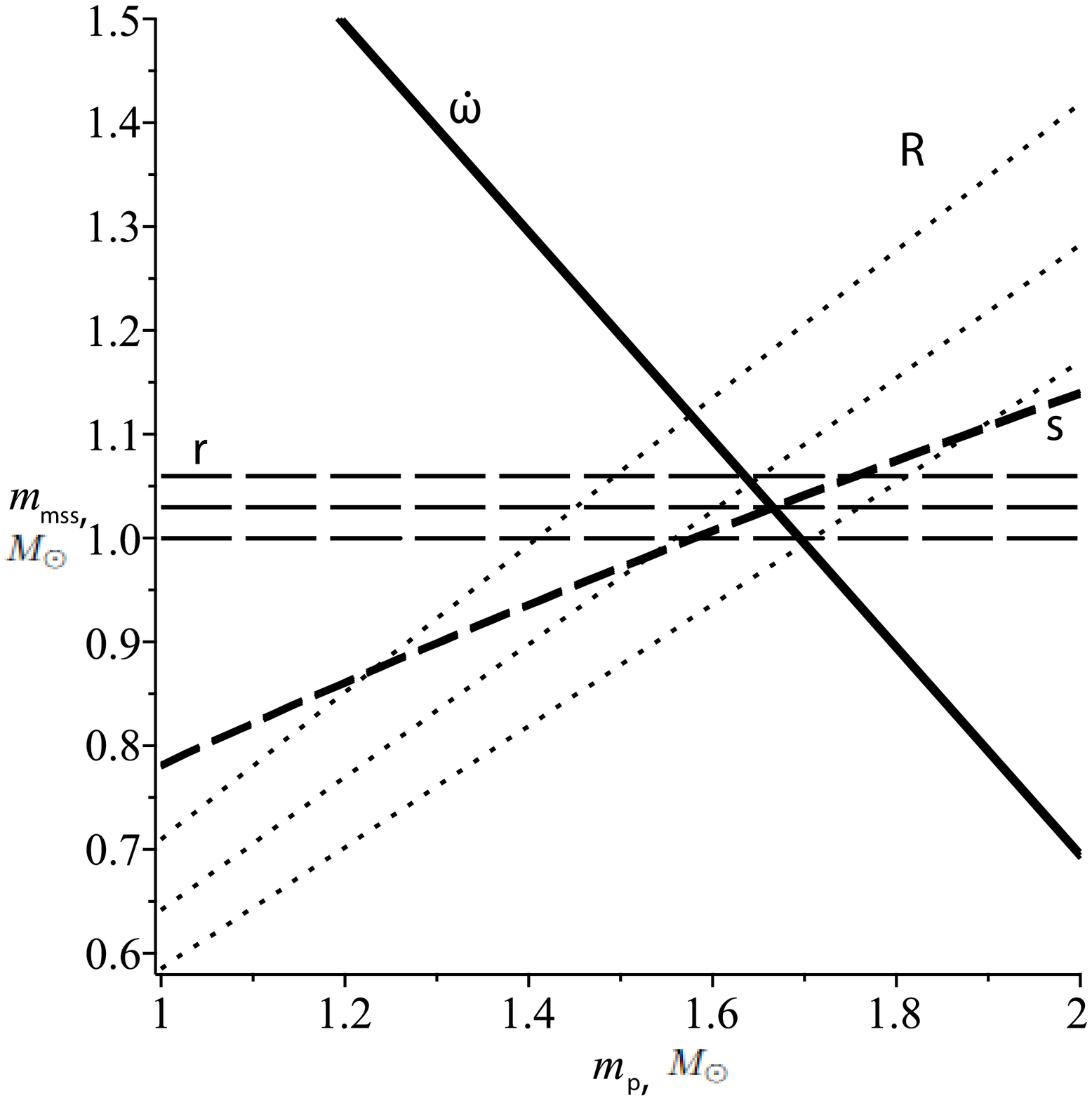}\label{pic GRm 1}}
		
		{b)\includegraphics[width=0.8\linewidth]{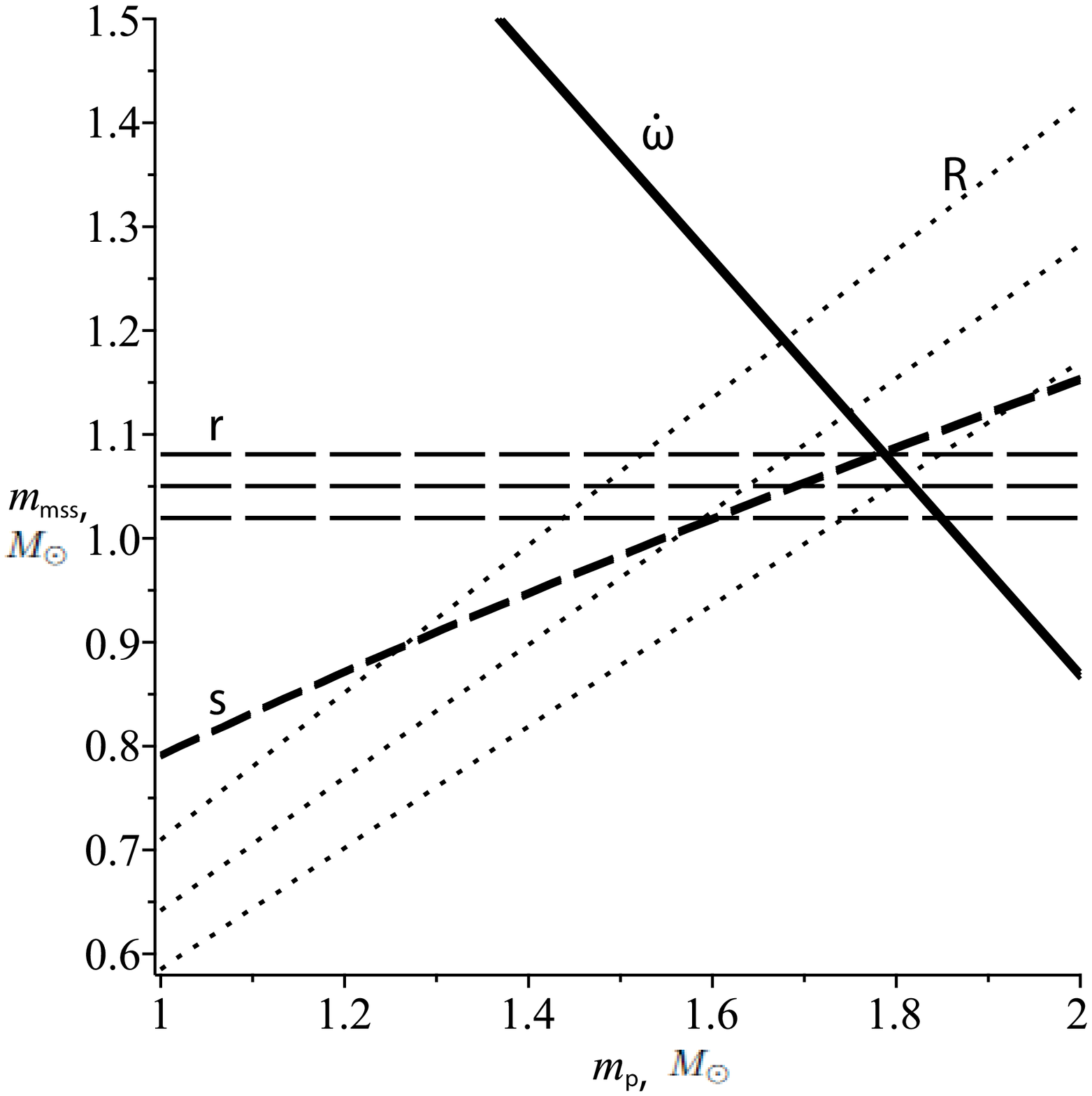}\label{pic GRm 2}}
		\caption{Mass-mass diagrams for PSR J1903+0327.
The width of each curve represents $\pm1\sigma$ error bounds.
a) GR case; b) hybrid f(R)-gravity case at boundary values of model parameters. }
		\label{pic GRm}
	\end{figure}

	As another system, we chose PSR J1903+0327, because in this binary pulsar three post-Keplerian parameters are measured quite accurately and the system has a large eccentricity.	Unfortunately in this system it is not possible to extract the part of orbital period change which is connected with gravitational radiation. Thus, we based on $\dot \omega, r, s$ and also on mass ratio which is measured in theory-independent way~\cite{freire}. Using data from PSR J1903+0327 (see Table \ref{1903}) we obtain this restrictions: 
$\phi_0 < 0.02$ and $\frac{\phi_0 m_{\phi}}{(1+\phi)^{\frac{1}{3}}(1-\frac{\phi_0}{3})^{\frac{2}{3}}}<2\times 10^{-20}$. Results of our investigation are presented on the fig. \ref{pic GRm}.
As we can see restrictions on $\phi_0$ from data of PSR J0737-3039 better than from data of PSR J1903+0327. In this system we also obtained predictions for the system components masses: $1.021 \ \  M_\odot<m_{\rm mss}<1.08 \ \  M_\odot$ and $1.646 \ \  M_\odot<m_{\rm p}<1.785 \ \  M_\odot$ ($m_{\rm p}$ is mass of the pulsar, $m_{\rm mss}$ is mass of its companion). The upper limit  will also exceed the upper limit predicted by GR: $1.021 \ \  M_\odot<m_{\rm mss}<1.037 \ \  M_\odot$ and $1.646 \ \  M_\odot<m_{\rm p}<1.688 \ \  M_\odot$~\cite{freire}.

\begin{table}[h]
\caption{Parameters  PSR J1903+0327}\label{1903}
\begin{tabular}{|l|l|l|}
\hline
Parameter & Physical & Experimental\\
 & meaning & value\\
\hline
$P_{\rm b} (day)$ & orbital period& $ 95.174118753(14)  $\\

\hline

$e$ & eccentricity & $ 0.436678409(3)   $\\
\hline

$x (lt-s)$ & projected   & $ 105.5934643(5)  $\\
& semimajor axis & \\
& of the pulsar orbit &\\

\hline

$\dot \omega(deg/yr)$ & secular   advance of & $ 0.0002400(2) $\\
&the periastron&\\

\hline
$s$ & Shapiro  ``shape'' & $ 0.9760(15) $\\
 & delay & \\
  &  parameter & \\
\hline
$r(\mu s)$ & Shapiro  ``range'' & $  1.03(3)  $\\
 & delay  & \\
  & parameter & \\
\hline
$R=\cfrac{m_{\rm p}}{m_{\rm mss}}$ & mass ratio & $ 1.55(20)  $\\
\hline
\end{tabular}
\end {table}

	\section{Conclusions}\label{sec:conclusions}
	In this work we developed parametrized post-Keplerian formalism for hybrid metric-Palatini f(R)-gravity. We obtained analytical expressions for four PPK parameters: $\dot \omega, \dot P_{\rm b}, r, s$. Unlike previous work~\cite{memnras}, we considered binary pulsars with eccentric orbits. To impose restrictions on the model parameters of hybrid f(R)-gravity we used the observational data of PSR J0737-3039 and PSR J1903+0327. We constrained the background value of the scalar field and the combination of parameters containing the scalar field mass.
	
	In addition we obtained the expression for orbital period change for binary pulsars with eccentric orbit. It includes tensor quadrupole and scalar monopole, monopole-quadrupole and quadrupole contributions. We showed that hybrid f(R)-gravity does not predict the existence of scalar dipole radiation.
	
	Also we obtained masses of components in systems PSR J0737-3039 and PSR J1903+0327 in the framework of hybrid f(R)-gravity. We showed that hybrid f(R)-gravity predicts larger masses than GR. This result was expected, as in~\cite{stars} it was shown that in the frameworks of hybrid f(R)-gravity neutron stars are heavier than in GR.
	
	The main purpose of this article was to test  hybrid f(R)-gravity with light scalar field using the observational data obtained from astrophysical objects with strong gravitational field like binary pulsars. Previously, we showed the viability of this model  testing it with only one PPK parameter the orbital period change  in  approximation of quasicircular orbits~\cite{memnras}. In this paper we provided more complete PPK test  in the more general case of eccentric orbit. Thus, this article continues a number of our works~\cite{memnras, meppn}, where we proved that the existence of a light scalar field in hybrid metric-Palatini f(R)-gravity that generates long-range forces does not contradict the data obtained from local systems with both a weak (Solar System) and a strong gravitational field (binary pulsars).
	
	As the next step, we plan to carry out a theoretical calculations of the mass-radius-luminosity dependence in the framework of hybrid f(R)-gravity and to impose restrictions on the free parameters of this theory from observational photometric data. Such study will also allow to obtain predictions of hybrid f(R)-theory regarding the masses of main sequence stars. Since the companion of the pulsar in the system  PSR J1903+0327 is a the main sequence star, it becomes possible to compare the mass of this object obtained from pulsar timing and from photometric data in the framework of hybrid f(R)-gravity, what will become an additional reliable test of this theory.

This work was supported by the grant 18-32-00785 from Russian Foundation for Basic Research.

\end{document}